\documentclass[twocolumn,showpacs,preprintnumbers,amsmath,amssymb,superscriptaddress,prb]{revtex4-2}
\usepackage{dcolumn}% Align table columns on decimal point
\usepackage{bm}% bold math
\usepackage[dvipdfmx,hidelinks]{hyperref}
\newcommand{\doijournal}[2]{\href{https://doi.org/#1}{#2}}
\usepackage[dvipdfmx]{graphicx}

\begin{document}

%\preprint{APS/123-QED}

\title{Optical vortex probe of loop-current chirality in moir\'e materials}

\author{Nobuhiko Yokoshi}
 \affiliation{Department of Physics and Electronics, Osaka Metropolitan University, 1-1 Gakuen-cho, Sakai, Osaka 599-8531, Japan}
\author{Akihito Kato}
 \affiliation{Department of Physics and Electronics, Osaka Metropolitan University, 1-1 Gakuen-cho, Sakai, Osaka 599-8531, Japan}
 \affiliation{Institute for Molecular Science, National Institutes of Natural Sciences, Okazaki 444-8585, Japan}

\date{\today}% It is always \today, today,
             %  but any date may be explicitly specified

\begin{abstract}
We propose a symmetry-resolved optical probe of intrinsic loop-current chirality in moir\'e materials, with twisted bilayer graphene as a representative realization. Interlayer interference generates chiral electronic circulation on triangular plaquettes, giving rise to an intrinsic geometric chirality that enters the second-order response through a $C_3$-selected angular harmonic of the Berry curvature and can be isolated by the orbital-angular-momentum difference $\Delta\ell$ of interfering optical vortex beams. When moir\'e $C_3$ symmetry is preserved, the intrinsic contribution appears in the $\Delta\ell=3$ channel of the helicity-dependent dc photocurrent, whereas $C_3$-breaking perturbations activate additional channels. These results establish angular-momentum-resolved nonlinear optics as a route to probing geometric chirality in moir\'e and other symmetry-engineered quantum materials.
\end{abstract}

\maketitle

%/_/_/_/_/_/_/_/_/_/_/_/_/_/_/_/_/_/_/_/_/_/_/_/_/_/_/_/_/_/_/_/_/_/_/_/_/_/_/_
\section{Introduction}
%/_/_/_/_/_/_/_/_/_/_/_/_/_/_/_/_/_/_/_/_/_/_/_/_/_/_/_/_/_/_/_/_/_/_/_/_/_/_/_

\begin{figure}[t]
  \centering
  \includegraphics[width=\linewidth]{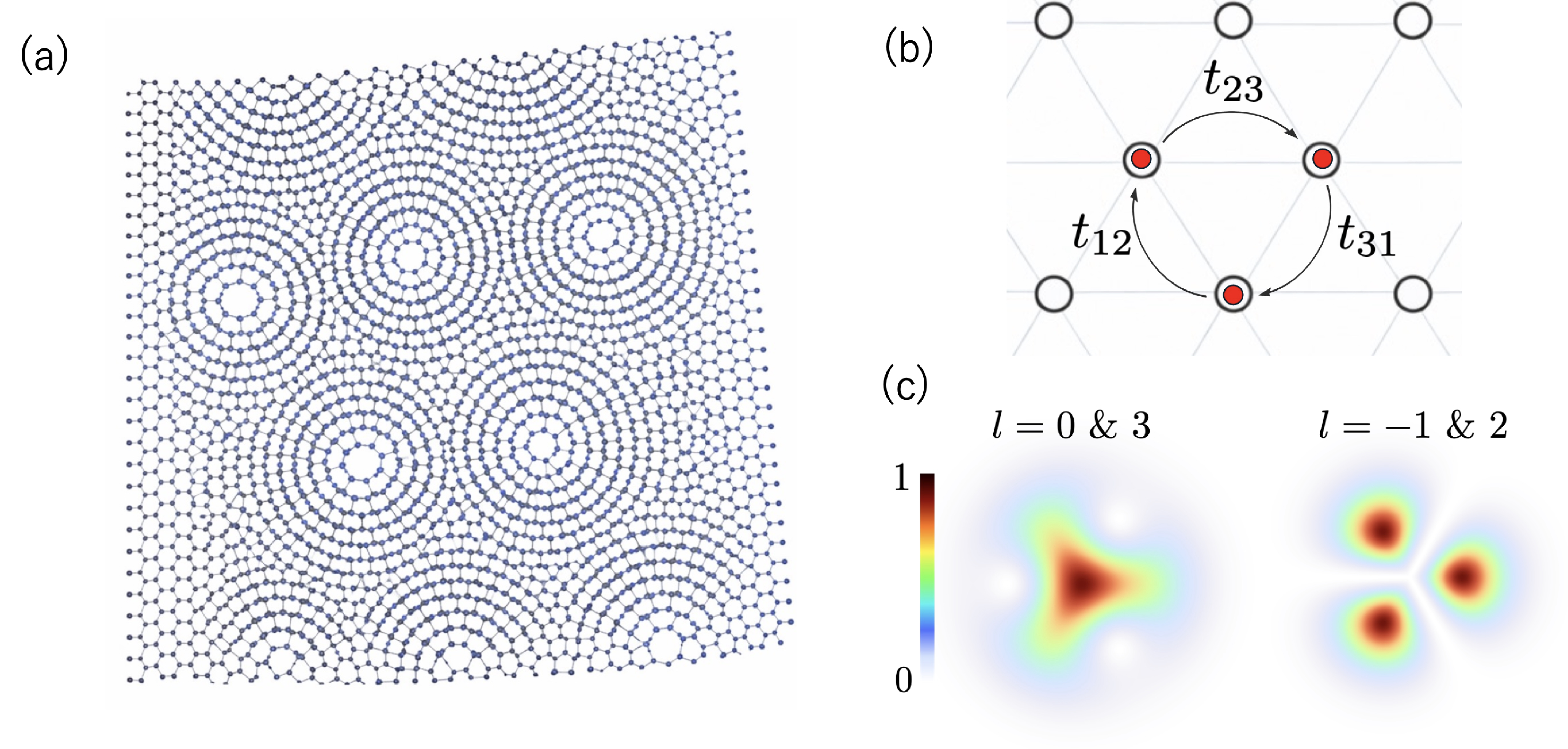}
\caption{
(a) Schematic of TBG, where a small twist angle between two graphene layers gives rise to a moir\'e superlattice.
(b) Three selected moir\'e-superlattice sites (filled red circles; open circles indicate neighboring sites) form an elementary triangular plaquette connected by effective interlayer-assisted hoppings $t_{12}$, $t_{23}$, and $t_{31}$.
(c) Intensity profiles of optical fields formed by superposing two optical vortices with different OAM values, exhibiting $C_3$ symmetry.
}
  \label{fig:fig1}
\end{figure}

Moir\'e materials provide a versatile setting for engineering electronic geometry beyond conventional band structures, with quantum interference rather than atomic-scale symmetry shaping electronic wave functions~\cite{Andrei2021NRM,Balents2020NatPhys,Sharpe2019Scienceaaw3780}.
In twisted bilayer graphene (TBG) and related systems, interlayer hybridization generates interference around triangular moir\'e plaquettes, producing circulating electronic motion and geometric phases tied to the moir\'e lattice symmetry [Fig.~\ref{fig:fig1}(a,b)]~\cite{Bistritzer2011PNAS,Cao2018CI,Cao2018SC,NamKoshino2017PRB,GuineaWalet2019PRB,KoshinoNam2020PRB,Po2018PRX}.
This quantum-mechanical circulation endows the moir\'e superlattice with an intrinsic chirality, reflecting the handedness of electronic motion on the triangular moir\'e network.
Its direct observation remains challenging, because conventional transport and optical probes generally average over the angular structure of the electronic response.

Nonlinear optical responses have become an important probe of geometric properties in quantum materials~\cite{NagaosaMorimoto2022AnnPhys,Mikhailov2007EPL,Cheng2014NJP}.
Second-order dc responses, including the circular photogalvanic effect and related rectification phenomena, are sensitive to Berry curvature and quantum geometry, and thus access geometric aspects of Bloch bands~\cite{Sipe2000PRB,MorimotoNagaosa2016PRB,Hosur2011PRB,deJuan2017NatCommun,Gao2020PRL,Farias2013EPJB}.
In their usual form, however, these probes integrate over angular components of the response kernel and therefore obscure the symmetry-resolved structure of intrinsic chirality.
Moir\'e materials are especially suitable in this context, because their long-wavelength superlattice potentials generate narrow bands with enhanced geometric effects and chiral electronic structure~\cite{SodemannFu2015PRL,Ma2019Nature,Kang2019NatMater}.

Structured light carrying orbital angular momentum (OAM) offers a way to resolve this angular structure~\cite{Allen1992PRA,YaoPadgett2011AOP}.
Because an optical vortex carries a definite azimuthal phase, it acts as an angular-momentum filter and couples selectively to angular harmonics that are absent in conventional optical probes~\cite{Quinteiro2010PRB,Watzel2016SciRep,HinoHiguchi2018JPSJ,Ishii2019JPCS_OV_TMD,Saito2021PCCP17271,Quinteiro2022RMP035003}.
Optical vortex beams therefore provide a natural route to symmetry-resolved geometric responses in moir\'e systems~\cite{Ishii2019JPCS_OV_TMD,Quinteiro2022RMP035003,HinoHiguchi2018JPSJ,Saito2021PCCP17271,Goto2021NJP_OAM_SpinSpin}.

In this paper, we show that optical vortex beams provide a symmetry-resolved nonlinear probe of intrinsic moir\'e chirality in TBG.
We consider coherent superpositions of vortex beams whose interference pattern is compatible with the underlying $C_3$ symmetry of the moir\'e lattice [Fig.~\ref{fig:fig1}(c)]~\cite{Zhao2025Nanophotonics4311}.
Within this framework, the intrinsic chiral contribution to the helicity-dependent dc photocurrent is selected by the OAM-difference channel $\Delta\ell=3$ as long as the moir\'e $C_3$ symmetry is preserved.
Symmetry-breaking perturbations activate additional OAM channels, providing a handle for separating intrinsic chirality from extrinsic anisotropy~\cite{Levy2010Science,Guinea2010NatPhys,Kogl2023npj2DM}.

%/_/_/_/_/_/_/_/_/_/_/_/_/_/_/_/_/_/_/_/_/_/_/_/_/_/_/_/_/_/_/_/_/_/_/_/_/_/_/_
\section{Loop-current chirality}
%/_/_/_/_/_/_/_/_/_/_/_/_/_/_/_/_/_/_/_/_/_/_/_/_/_/_/_/_/_/_/_/_/_/_/_/_/_/_/_

At small twist angles, TBG is well described by the Bistritzer--MacDonald (BM) continuum moir\'e Hamiltonian~\cite{Bistritzer2011PNAS},
\begin{equation}
H_\xi
=
\begin{pmatrix}
h_{\xi}^{(1)}(\bm k) & T_\xi(\bm r) \\
T_\xi^\dagger(\bm r) & h_{\xi}^{(2)}(\bm k)
\end{pmatrix},
\label{eq:BM_Hamiltonian}
\end{equation}
where $\xi=\pm$ labels the two valleys, $h_{\xi}^{(l)}(\bm k)$ is the Dirac Hamiltonian of layer $l$ rotated by $\pm\theta/2$, and $T_\xi(\bm r)$ is the moir\'e-periodic interlayer tunneling matrix; in the optical response, the field couples through the velocity operator $v_a=(1/\hbar)\partial H_\xi/\partial k_a$ with $a=x,y$.
The tunneling term is expanded as
\begin{equation}
T_\xi(\bm r)
=
\sum_{j=1}^{3}
T_j^{(\xi)} e^{i\xi \bm q_j\cdot \bm r},
\label{eq:BM_tunneling}
\end{equation}
The associated moir\'e wave vectors form a $C_3$-related triad, $\bm q_j=k_\theta(\cos\phi_j,\sin\phi_j)$ with $\phi_j=0,2\pi/3,4\pi/3$ and $k_\theta=\frac{8\pi}{3a}\sin\!\left(\frac{\theta}{2}\right)$,
where $a=0.246~\mathrm{nm}$ is the graphene lattice constant~\cite{Bistritzer2011PNAS}.
The relative phases of $T_j^{(\xi)}$ encode the stacking-dependent interlayer hybridization and define the minimal three-wave moir\'e interference structure.
This continuum description is most reliable for small twist angles, typically $\theta\lesssim 2^\circ$, and often remains qualitatively useful up to a few degrees as long as higher-harmonic scattering and states away from the Dirac points remain subdominant~\cite{NamKoshino2017PRB,GuineaWalet2019PRB,KoshinoNam2020PRB}.

In the same BM framework, the intrinsic chirality is associated with a geometric phase accumulated around an elementary triangular moir\'e plaquette.
For three effective interlayer-assisted hoppings $t_{12}$, $t_{23}$, and $t_{31}$ connecting the sites of the plaquette [Fig.~\ref{fig:fig1}(b)], we define
\begin{equation}
\Phi_\triangle = \arg(t_{12}t_{23}t_{31}),
\label{eq:loop_phase}
\end{equation}
the phase of the closed-loop hopping product.
Because this quantity is unchanged by a local rephasing of the moir\'e-site basis, it is a gauge-invariant geometric phase~\cite{Haldane1988PRL2015}.
Reversing the circulation direction complex-conjugates the product $t_{12}t_{23}t_{31}$ and hence changes the sign of $\Phi_\triangle$, showing that it is a pseudoscalar measure of the handedness of the triangular interference process.
Its twist-angle dependence is constrained by symmetry: $\Phi_\triangle$ must be odd under $\theta\to-\theta$ and must vanish at $\theta=0$, where no intrinsic moir\'e chirality is generated.
Accordingly, the leading $C_3$-allowed dependence scales as $\Phi_\triangle\sim\theta^3$ for small $\theta$; a more explicit form and its derivation are given in the Appendix~Sec.\ref{appAa}.
At larger twist angles, higher-harmonic moir\'e couplings can renormalize the detailed form of $\Phi_\triangle$ without changing its geometric or pseudoscalar character.

The same chirality can be described in momentum space by the Berry curvature of the moir\'e bands~\cite{Xiao2010RMP,Nagaosa2010RMP}.
When the valley index remains a good quantum number, the relevant quantity is the valley-contrasting combination
$\Omega_z(\bm k)=\sum_{\xi=\pm}\xi\,\Omega_z^{\xi}(\bm k)$, where $\xi=\pm$ labels the two valleys~\cite{Mak2012NatNano,Xu2014NatPhys,Xiao2010RMP}.
In TBG, the twist produces spatially varying local stacking and thereby imprints complex phases on the interlayer tunneling, which endow the Berry curvature with a pronounced angular dependence~\cite{Bistritzer2011PNAS,NamKoshino2017PRB,GuineaWalet2019PRB,KoshinoNam2020PRB,Po2018PRX}.
Because the moir\'e pattern is triangular, the corresponding intrinsic chirality is captured by the $m=3$ angular harmonic
\begin{equation}
\Omega^{(3)} \equiv
\sum_{\xi=\pm}\xi\int_{\rm BZ}\frac{d^2k}{(2\pi)^2}\,
\Omega_z^{\xi}(\bm k)\,e^{i3\phi_k},
\label{eq:Omega3}
\end{equation}
writing $\bm k=k(\cos\phi_k,\sin\phi_k)$.
This quantity isolates the lowest nontrivial angular component tied to the triangular moir\'e geometry.
If additional internal degrees of freedom are present, the same definition applies after summing over the corresponding quantum numbers within each valley.
Although $\Omega^{(3)}$ may be computed from a microscopic band-structure model, the discussion below relies only on its symmetry content and on the fact that it vanishes when the moir\'e-induced $C_3$ interference is absent~\cite{Bistritzer2011PNAS}.

A complementary real-space characterization is obtained by projecting the BM continuum Hamiltonian onto an effective moir\'e lattice description.
In this representation, the interlayer-assisted tunneling processes encoded in Eq.~\eqref{eq:BM_tunneling} generate effective hoppings $t_{ij}$ between neighboring moir\'e sites, from which the directed bond-current operator $\hat J_{ij}$ is defined (see the Appendix~Sec.\ref{appAb}).
We then introduce the loop-circulation operator on an elementary triangular plaquette $\langle ijk\rangle$, $\hat O_{ijk}\equiv \hat J_{ij}+\hat J_{jk}+\hat J_{ki}$, which sums the bond currents around the plaquette.
Because $\hat O_{ijk}$ changes sign under reversal of the loop orientation, it provides a natural measure of local circulation.
The intrinsic chirality is then quantified by the pseudoscalar
\begin{equation}
\chi \equiv \frac{1}{N_\triangle}\sum_{\langle ijk\rangle}
\langle \hat O_{ijk}-\hat O_{ikj} \rangle ,
\label{eq:intrinsic_chirality}
\end{equation}
with $N_\triangle$ the number of elementary triangular plaquettes included in the sum.
A related notion of loop-current order and its chiral extension has been discussed in cuprates~\cite{Varma2014JPCM505701,Pershoguba2013PRL047005}.
Within the minimal three-wave picture, the loop phase controls the sign and magnitude of this quantity, yielding $\chi\propto\sin\Phi_\triangle$ to leading order.

This chirality should be distinguished from a static orbital magnetization.
It can remain well defined even in equilibrium, but when time-reversal symmetry is preserved it is valley contrasting, so the net chirality vanishes for equal valley occupation~\cite{Xiao2010RMP,Nagaosa2010RMP}.
The quantities $\Phi_\triangle$, $\Omega^{(3)}$, and $\chi$ therefore provide complementary descriptions of the same intrinsic moir\'e chirality probed by the nonlinear optical response discussed below.

%/_/_/_/_/_/_/_/_/_/_/_/_/_/_/_/_/_/_/_/_/_/_/_/_/_/_/_/_/_/_/_/_/_/_/_/_/_/_/_
\section{Optical vortex probe}
%/_/_/_/_/_/_/_/_/_/_/_/_/_/_/_/_/_/_/_/_/_/_/_/_/_/_/_/_/_/_/_/_/_/_/_/_/_/_/_

We now formulate the optical-vortex probe of loop-current chirality.
The central observation is that the nonlinear optical response inherits the same three-wave moir\'e structure as the BM Hamiltonian, while the OAM of light serves to filter its angular harmonics.
In this way, the intrinsic $C_3$-symmetric chirality of the moir\'e system is brought into direct correspondence with selected OAM channels in the helicity-dependent dc photocurrent.

From a microscopic point of view, the nonlinear response is built from virtual optical transitions dressed by the three moir\'e scattering channels $\bm q_j$.
A representative amplitude in channel $j$ is
\begin{equation}
\mathcal M_{\omega}^{(j)}(\bm k)
\sim
\sum_n
\frac{
\langle u_{\bm k}|v_a|u_{\bm k,n}\rangle\,
\langle u_{\bm k,n}|\mathcal T_j|u_{\bm k+\bm q_j}\rangle
}{
\omega-(\varepsilon_{\bm k,n}-\varepsilon_{\bm k})+i0^+
},
\label{eq:Ma_schematic}
\end{equation}
where $|u_{\bm k}\rangle$ is the initial Bloch state, $|u_{\bm k,n}\rangle$ is an intermediate state, and $\mathcal T_j$ denotes scattering through the $j$th moir\'e channel.
Because the three $\bm q_j$ form a $C_3$ triplet, these amplitudes are related by
$\mathcal M_{\omega}^{(j)}(R_{2\pi/3}\bm k)=\mathcal M_{\omega}^{(j+1)}(\bm k)$,
with $j \ {\rm mod}\ 3$, where $R_{2\pi/3}$ denotes the in-plane rotation by $2\pi/3$ about the $z$ axis (see the Appendix~Sec.\ref{appAc}).

The dc response kernel $\mathcal K_\omega(\bm k)$ is obtained by combining these three channel contributions and their vertex permutations.
Accordingly, it admits the schematic decomposition
\begin{equation}
\mathcal K_{\omega}(\bm k)
=
\sum_{j=1}^{3}
\mathcal F_{\omega}^{(j)}(\bm k)\,
e^{i(\bm k \cdot \bm q_j)/k_\theta^2}
=
\sum_m
\mathcal K_{\omega}^{(m)}(\bm k)e^{im\phi_k},
\label{eq:kernel_3channel}
\end{equation}
with channel weights $\mathcal F_{\omega}^{(j)}$ inheriting the same cyclic $C_3$ relation.
Therefore a $C_3$-symmetric scalar kernel contains only harmonics $m=3n$.
Using
$e^{i(\bm k\cdot\bm q_j)/k_\theta^2}=e^{i(k/k_\theta)\cos(\phi_k-\phi_j)}$
and
$\sum_{j=1}^{3}e^{-im\phi_j}=3\,\delta_{m\equiv0\ ({\rm mod}\ 3)}$,
one finds
$\mathcal K_{\omega}^{(m)}(\bm k)\neq0$
only for
$m=0,\pm3,\pm6,\ldots$.
The lowest nontrivial angular sector is thus $m=\pm3$, which gives the leading intrinsic channel tied to the triangular moir\'e geometry.

We next introduce the optical field.
An optical vortex with OAM $\ell$ carries an azimuthal phase factor $e^{i\ell\phi}$~\cite{Allen1992PRA,YaoPadgett2011AOP,Quinteiro2010PRB,Watzel2016SciRep,Quinteiro2022RMP035003}.
For a coherent superposition of two vortex beams with OAMs $\ell_1$ and $\ell_2$,
\begin{equation}
\bm E(\bm r,\omega)
=
\bm E_{\ell_1}(\bm r)e^{i\ell_1\phi}
+
\bm E_{\ell_2}(\bm r)e^{i\ell_2\phi},
\label{eq:E_superposition}
\end{equation}
the interference term carries the angular factor
$e^{i(\ell_1-\ell_2)\phi}\equiv e^{i\Delta\ell\phi}$.
This allows the nonlinear response to be projected onto a selected angular harmonic.
We therefore define the OAM-resolved spectral weight
\begin{equation}
\begin{split}
\mathcal A^{(\Delta\ell)}(\omega)
&=
\int_{\rm BZ}\frac{d^2k}{(2\pi)^2}\,
e^{i\Delta\ell\phi_k}\,
\mathcal K_{\omega}(\bm k)\,\Gamma_\omega(k)\\
&\equiv
\Bigl\langle
e^{i\Delta\ell\phi_k}\,
\mathcal K_{\omega}(\bm k)
\Bigr\rangle_{\omega},
\end{split}
\label{eq:Aell}
\end{equation}
where $\Gamma_\omega(k)$ is a smooth resonance weight incorporating the optical transition condition, linewidth, and occupation factors~\cite{Sipe2000PRB}.
The dc photocurrent can then be decomposed as
\begin{equation}
j_{\omega}^{(2)}(0)=\sum_{\Delta\ell}j_{\omega}^{(\Delta\ell)}(0),
~~
j_{\omega}^{(\Delta\ell)}(0)=\mathcal P(\omega)\,\mathcal A^{(\Delta\ell)}(\omega),
\label{eq:jDeltaell}
\end{equation}
with $\mathcal P(\omega)\propto\int d^2r\,[\bm E_{\ell_1}(\bm r)\!\cdot\!\bm E_{\ell_2}^\ast(\bm r)+{\rm c.c.}]$ determined by the beam profiles.

The selection rule follows directly from the angular decomposition in Eq.~\eqref{eq:kernel_3channel}.
Substituting it into Eq.~\eqref{eq:Aell}, one obtains
\begin{equation}
\mathcal A^{(\Delta\ell)}(\omega)
=
\sum_m
\int_{\rm BZ}\frac{d^2k}{(2\pi)^2}\,
\Gamma_\omega(k)\,
\mathcal K_{\omega}^{(m)}(\bm k)\,
e^{i(m+\Delta\ell)\phi_k}.
\label{eq:selection_rule}
\end{equation}
For an angularly isotropic resonance weight and within the circularized continuum approximation to the moir\'e Brillouin zone, the angular integral projects the response onto the $m=-\Delta\ell$ sector.
Thus, the OAM difference $\Delta\ell$ probes the $m=-\Delta\ell$ angular sector of the intrinsic response kernel.
Since a $C_3$-symmetric kernel contains only $m=3n$ sectors, only $\Delta\ell=3n$ channels contribute to the intrinsic response.
The lowest nontrivial intrinsic signal is therefore selected by $|\Delta\ell|=3$, as in Fig.~\ref{fig:fig1}(c).

This angular selection rule also clarifies the origin of the chiral $\Delta\ell=3$ response.
In Eq.~\eqref{eq:kernel_3channel}, the nonlinear kernel is decomposed into three channel-resolved contributions $\mathcal F_{\omega}^{(j)}(\bm k)$ associated with the three moir\'e scattering channels.
Equation~\eqref{eq:Ma_schematic} extracts from each $\mathcal F_{\omega}^{(j)}$ a representative microscopic amplitude $\mathcal M_{\omega}^{(j)}(\bm k)$ carrying the same channel-dependent phase information.
A single channel amplitude can affect the overall magnitude of the response, but it cannot distinguish the sense of circulation on the moir\'e triangle.
The handedness is encoded only in the relative phase accumulated across the full $C_3$-related channel triplet.
The lowest gauge-invariant quantity is therefore the closed product
\begin{equation}
\mathcal I_\triangle(\bm k,\omega)
\equiv
\mathcal M_{\omega}^{(1)}(\bm k)\,
\mathcal M_{\omega}^{(2)}(\bm k)\,
\mathcal M_{\omega}^{(3)}(\bm k),
\label{eq:Itriangle}
\end{equation}
which is the optical analog of the real-space loop product $t_{12}t_{23}t_{31}$ introduced in Eq.~\eqref{eq:loop_phase}.
If one triangular orientation is represented by the ordered sequence
$1\!\to\!2\!\to\!3\!\to\!1$,
the reversed orientation
$1\!\to\!3\!\to\!2\!\to\!1$
is represented by $\mathcal I_\triangle^\ast$.
Accordingly, the orientation-even part is $\mathrm{Re}\,\mathcal I_\triangle$, whereas the orientation-odd, namely chiral, part is $\mathrm{Im}\,\mathcal I_\triangle$~\cite{KatoYokoshi2024PRB_HHG_Chirality}.

After the OAM projection has selected the $m=-3$ sector, the $\Delta\ell=3$ signal is determined by the orientation-odd part of this closed three-channel contribution, so that
\begin{equation}
\mathcal A_{\rm chiral}^{(\Delta\ell=3)}(\omega)
\propto
\Bigl\langle
{\rm Im}\!\left[
\mathcal M_{\omega}^{(1)}(\bm k)\,
\mathcal M_{\omega}^{(2)}(\bm k)\,
\mathcal M_{\omega}^{(3)}(\bm k)
\right]
\Bigr\rangle_{\omega}.
\label{eq:Achiral3}
\end{equation}
Writing
$\mathcal I_\triangle(\bm k,\omega)=|\mathcal I_\triangle(\bm k,\omega)|e^{i\Phi_\triangle(\bm k,\omega)}$,
one has
${\rm Im}\,\mathcal I_\triangle=|\mathcal I_\triangle|\sin\Phi_\triangle(\bm k,\omega)$.
Equation~\eqref{eq:Achiral3} therefore shows that the chiral optical signal is nonzero only when the closed three-channel virtual process accumulates a finite loop phase.
It is then natural to write the helicity-odd contribution in the factorized form
\begin{equation}
\mathcal A_{\rm chiral}^{(\Delta\ell=3)}(\omega)
\equiv
\chi\,\tilde{\mathcal A}_{\rm chiral}^{(3)}(\omega),
\label{eq:Achi_factorized}
\end{equation}
where $\tilde{\mathcal A}_{\rm chiral}^{(3)}(\omega)$ contains the remaining frequency-dependent microscopic weight.

The same symmetry framework also identifies how additional OAM channels are activated by extrinsic symmetry breaking.
A representative case is lattice strain, which breaks $C_3$ and introduces an $m=\pm2$ angular component into the nonlinear kernel~\cite{Levy2010Science,Guinea2010NatPhys,Kogl2023npj2DM}.
Writing the symmetry-breaking part of the strain tensor as
$\varepsilon_2=(\varepsilon_{xx}-\varepsilon_{yy})+2i\varepsilon_{xy}
=\varepsilon_\ast e^{i2\phi_\varepsilon}$,
the leading strain-induced correction takes the form
\begin{equation}
\delta\mathcal K_{\omega}(\bm k)
=
\varepsilon_2^\ast\,\mathcal Q_{\omega}(k)\,e^{i2\phi_k}
+
\varepsilon_2\,\mathcal Q_{\omega}^\ast(k)\,e^{-i2\phi_k},
\label{eq:deltaK_strain}
\end{equation}
where $\mathcal Q_{\omega}(k)$ is a strain-induced response function determined by the band structure and interlayer tunneling~\cite{NamKoshino2017PRB,GuineaWalet2019PRB,KoshinoNam2020PRB}.
Equation~\eqref{eq:selection_rule} then implies a response in the $\Delta\ell=\pm2$ channels.
We therefore write
\begin{equation}
\mathcal A^{(\Delta\ell=2)}(\omega)
\equiv
\varepsilon_2^\ast\,\tilde{\mathcal A}^{(2)}(\omega),
\label{eq:A2strain}
\end{equation}
so that the ratio of the intrinsic $\Delta\ell=3$ channel and the strain-induced $\Delta\ell=2$ channel removes the common geometric factor $\mathcal P(\omega)$,
\begin{equation}
R(\omega)
=
\frac{j_{\omega,{\rm chiral}}^{(\Delta\ell=3)}}
     {j_{\omega}^{(\Delta\ell=2)}}
=
\frac{\chi}{\varepsilon_2^\ast}
\frac{\tilde{\mathcal A}_{\rm chiral}^{(3)}(\omega)}
     {\tilde{\mathcal A}^{(2)}(\omega)}.
\label{eq:Rratio}
\end{equation}
Here $j_{\omega}^{(\Delta\ell)}$ denotes the complex amplitude of the $\Delta\ell$ channel in the phase-harmonic decomposition, rather than the directly observed current itself.
Equation~\eqref{eq:Rratio} therefore defines a ratio of channel amplitudes: its magnitude provides a per-device measure of the intrinsic chiral response relative to strain at fixed twist angle, while its phase reflects the orientation of the principal strain axis.
More generally, perturbations transforming as $m=\pm2$ under in-plane rotations activate the same class of additional OAM channels, allowing intrinsic and extrinsic contributions to be distinguished within the same measurement.

The chiral part can be isolated experimentally by phase-sensitive detection of the interference between the two vortex-beam components.
Introducing a controllable relative phase $\delta$, the helicity-dependent dc signal is
\begin{equation}
j_{\omega}^{\rm int}(\delta)
\propto
2\,{\rm Re}\!\left[
\sum_{\Delta\ell}
j_{\omega}^{(\Delta\ell)}e^{i\Delta\ell\delta}
\right],
\label{eq:jint}
\end{equation}
which is manifestly real.
Thus each OAM channel appears as a distinct phase harmonic in $\delta$.
If the phase is modulated as $\delta(t)=\nu t$, the interference signal oscillates at harmonics $\Delta\ell\,\nu$, so that the intrinsic and strain-induced responses are isolated at $3\nu$ and $2\nu$, respectively.
Writing
$j_{\omega}^{(3)}=j_{\omega,{\rm achiral}}^{(3)}+i\,j_{\omega,{\rm chiral}}^{(3)}$,
the $\Delta\ell=3$ contribution contains an in-phase term proportional to $j_{\omega,{\rm achiral}}^{(3)}\cos(3\delta)$ and a quadrature term proportional to $j_{\omega,{\rm chiral}}^{(3)}\sin(3\delta)$.
Quadrature lock-in detection at $3\nu$ isolates the intrinsic chiral dc current, whereas lock-in detection at $2\nu$ yields the strain-induced $\Delta\ell=2$ response.
Comparing these measured lock-in amplitudes provides an experimental estimate of the intrinsic chiral response relative to strain.

A finite beam waist does not alter the local angular-momentum selection rule, but it reduces the absolute magnitude of the measured signal.
For an optical vortex beam with waist $w_0$, the illumination is localized by a smooth envelope, so that the measured response is a spatially averaged nonlinear current density~\cite{YaoPadgett2011AOP,Quinteiro2010PRB,Quinteiro2022RMP035003}.
In the regime $w_0\gg L_{\rm M}$, with $L_{\rm M}=a/[2\sin(\theta/2)]$, we write this finite-waist reduction as $j_{\omega}^{(\Delta\ell)}\simeq \eta_{\rm bw}j_{\omega,0}^{(\Delta\ell)}$, where $\eta_{\rm bw}\sim(L_{\rm M}/w_0)^2$.
For $L_{\rm M}\sim10$--$15~{\rm nm}$ and $w_0\sim1~\mu{\rm m}$, $\eta_{\rm bw}$ is of order $10^{-4}$.
This factor affects the signal magnitude but does not mix different $\Delta\ell$ channels.

In realistic TBG devices, an additional reduction may arise from rotational disorder of the moir\'e domains.
Related local twist-angle variations and strain inhomogeneity have been observed in Raman, low-energy electron diffraction, and spatially resolved microscopy measurements and discussed theoretically as sources of moir\'e disorder~\cite{Beechem2014ACSNano1655,Kazmierczak2021NatMater956,deJong2022NatCommun70,Wilson2020PRR023325}.
The selection rule is local and assumes a fixed in-plane orientation of the moir\'e pattern within the probed region.
Let $\alpha({\bf r})$ denote the local rotation angle of a chosen moir\'e axis, for example the direction of a reciprocal vector ${\bf q}_1({\bf r})$, measured relative to the optical interference pattern or to a fixed laboratory axis.
Because of $C_3$ symmetry, $\alpha$ is defined modulo $2\pi/3$, and the local orientation enters the chiral $\Delta\ell=3$ amplitude through $e^{i3\alpha({\bf r})}$.

When the optical spot covers several locally rotated domains, the measured chiral amplitude is the coherent spatial sum
\begin{equation}
j_{\omega,{\rm obs}}^{(3)}
=
\eta_{\rm bw} j_{\omega,0}^{(3)}{\cal D}_3,
\qquad
{\cal D}_3
=
\frac{\int d^2r\,W({\bf r})e^{i3\alpha({\bf r})}}
{\int d^2r\,W({\bf r})}.
\end{equation}
Here $W({\bf r})$ is a real, non-negative weight specifying the relative spatial contribution of each local region to the normalized orientational average ${\cal D}_3$.  It may be written, for example, as
$W({\bf r})\simeq I_{\ell_1\ell_2}({\bf r})|a_3({\bf r})|$.
Here $I_{\ell_1\ell_2}({\bf r})$ is the local overlap of the two vortex components that generate the $\Delta\ell=3$ interference term, while $a_3({\bf r})$ denotes the local complex amplitude of the intrinsic $\Delta\ell=3$ nonlinear response before orientational averaging.  Equivalently, $a_3({\bf r})$ represents the local strength of the chiral response associated with the three-channel loop process, including possible spatial variations of twist angle, carrier density, displacement field, relaxation, or local optical resonance conditions.  The orientational phase associated with the local moir\'e axis is not included in $a_3({\bf r})$ but is kept explicitly as $e^{i3\alpha({\bf r})}$ in Eq.~(19).  The overall finite-waist reduction of the signal magnitude is already included in $\eta_{\rm bw}$; therefore $W({\bf r})$ should be understood only as a relative weighting factor within ${\cal D}_3$.

For a single orientationally coherent domain, ${\cal D}_3=e^{i3\alpha}$ and $|{\cal D}_3|=1$.
For $N_{\rm eff}$ independently oriented domains, the factors $e^{i3\alpha}$ add as random phasors, giving $\langle{\cal D}_3\rangle=0$ and $|{\cal D}_3|_{\rm rms}\sim N_{\rm eff}^{-1/2}$.
With $N_{\rm eff}\sim\pi w_0^2/\xi_{\rm rot}^2$, where $\xi_{\rm rot}$ is the orientational correlation length, this gives $|{\cal D}_3|_{\rm rms} \sim \xi_{\rm rot}/(\sqrt{\pi}w_0)$.
Thus, for $w_0\sim1~\mu{\rm m}$ and $\xi_{\rm rot}\sim10$--$100~{\rm nm}$, rotational disorder can suppress the coherent $\Delta\ell=3$ signal by an additional factor of order $10^{-2}$--$10^{-1}$.
This orientational-dephasing factor is distinct from $\eta_{\rm bw}$: the former describes phase cancellation among locally rotated moir\'e domains, whereas the latter reflects finite-waist spatial averaging even for a fixed moir\'e orientation.

If the moir\'e orientation varies smoothly rather than forming randomly oriented domains, the suppression can be weaker.
For Gaussian fluctuations $\alpha({\bf r})=\bar{\alpha}+\delta\alpha({\bf r})$ with variance $\sigma_\alpha^2$, one obtains ${\cal D}_3=e^{i3\bar{\alpha}}e^{-9\sigma_\alpha^2/2}$.
The proposed $\Delta\ell=3$ probe is most favorable in devices with large orientationally coherent regions, or in scanning and near-field geometries where the effective optical area is reduced.
Such nanoscale photocurrent approaches have already resolved local electronic structure and photoresponse in TBG at moir\'e length scales~\cite{Sunku2020NanoLett2958,Hesp2021NatCommun1640}, supporting the feasibility of probing locally aligned regions.
Rotational disorder therefore reduces the coherent amplitude but does not invalidate the local $C_3$ selection rule.

Using typical graphene-based responsivities, $R_{\rm ph}\sim10$--$20~{\rm mA/W}$, the ideal single-domain estimate under $P_{\rm in}\sim1~{\rm mW}$ illumination gives a current scale of order $10$--$20~\mu{\rm A}$ before the finite-waist and orientational reductions are applied~\cite{Furchi2012NanoLett,Koppens2014NatNano}.
The experimentally observed chiral current should therefore be viewed as this local current scale multiplied by $\eta_{\rm bw}$, $|{\cal D}_3|$, and the microscopic weight of the $\Delta\ell=3$ nonlinear channel.
The resulting localized dc signal may be detected either as a terminal photocurrent or as a photovoltage after spatial integration with a suitable contact geometry~\cite{Ma2023NatRevPhys,Prechtel2012NatCommun}.
The phase-sensitive protocol remains useful in this reduced-signal regime because it separates the $\Delta\ell=3$ response from achiral backgrounds and from the strain-induced $\Delta\ell=2$ channel by their different phase harmonics.
The simultaneous measurement of the $\Delta\ell=2$ channel can also provide an in situ diagnostic of local strain and orientational inhomogeneity.

Because changing the twist angle $\theta$ reconstructs the moir\'e bands and modifies the optical matrix elements, an absolute comparison of the $\Delta\ell=3$ photocurrent across different twist angles does not by itself isolate the intrinsic chirality.
A more direct strategy is to work at fixed $\theta$ within a single device and tune control parameters such as carrier density or interlayer displacement field.
Within such a same-device protocol, comparing the extracted $3\nu$ and $2\nu$ lock-in amplitudes in the same resonance window reduces sample-to-sample ambiguity and helps separate the symmetry-selected chiral contribution from the overall band-structure dependence of the optical weight.

%/_/_/_/_/_/_/_/_/_/_/_/_/_/_/_/_/_/_/_/_/_/_/_/_/_/_/_/_/_/_/_/_/_/_/_/_/_/_/_
\section{Conclusion}
%/_/_/_/_/_/_/_/_/_/_/_/_/_/_/_/_/_/_/_/_/_/_/_/_/_/_/_/_/_/_/_/_/_/_/_/_/_/_/_

In this work, we have shown that a superposition of optical vortex beams provides a symmetry-resolved probe of intrinsic moir\'e loop-current chirality~\cite{Allen1992PRA,YaoPadgett2011AOP}.
In a moir\'e system preserving $C_3$ symmetry, the intrinsic helicity-dependent dc response is selected by the $\Delta\ell=3$ channel, giving optical access to a component that is averaged out in conventional measurements.
This selection rule follows from the threefold symmetry of the moir\'e interference process and remains valid as long as the underlying $C_3$ symmetry is preserved, although the microscopic response amplitudes can acquire quantitative twist-angle dependence beyond the minimal continuum description~\cite{Bistritzer2011PNAS,NamKoshino2017PRB,GuineaWalet2019PRB,KoshinoNam2020PRB,Po2018PRX}.
For finite beam waist, the response remains spatially localized and may be detected as a photocurrent or photovoltage after spatial integration, consistent with micrometer-scale focusing~\cite{YaoPadgett2011AOP,Quinteiro2022RMP035003}.
In realistic multi-domain samples, finite-waist averaging and orientational dephasing can substantially reduce the coherent $\Delta\ell=3$ amplitude, but these effects do not invalidate the local angular-momentum selection rule.
They instead point to concrete measurement strategies, such as using large orientationally coherent regions or scanning and near-field photocurrent geometries that reduce the effective optical area and minimize the number of independently oriented domains.

TBG provides a natural setting for this mechanism.
In dual-gate devices, an interlayer displacement field offers an additional handle for tuning the Berry curvature generated by moir\'e interference~\cite{Huang2023PRL_NLHE_GateSwitchable}.
Because such a field preserves moir\'e $C_3$ symmetry while breaking layer inversion, the intrinsic response remains associated with the $\Delta\ell=3$ channel, and its gate dependence can provide further information on the underlying chiral electronic structure~\cite{Huang2023PRL_NLHE_GateSwitchable}.
By contrast, weak $C_3$-breaking perturbations such as strain or substrate anisotropy activate additional OAM channels, allowing intrinsic and extrinsic contributions to be distinguished within the same measurement~\cite{Levy2010Science,Guinea2010NatPhys,Kogl2023npj2DM}.

More generally, the relation between loop-current chirality and a symmetry-selected angular harmonic is rooted in rotational symmetry and quantum interference, and should therefore extend beyond TBG to other moir\'e quantum materials~\cite{Dean2013Nature12186,Ponomarenko2013Nature12187,Du2023Scienceeadg0014}.
Although microscopic ingredients such as spin--orbit coupling, substrate potentials, and layer asymmetries can modify the band structure and optical matrix elements, the symmetry-based selection rule remains intact as long as the relevant rotational symmetry is preserved~\cite{Sipe2000PRB,Xiao2010RMP}.
OAM-resolved nonlinear optics therefore provides a general route to identifying intrinsic chiral responses hidden in conventional probes, while tracking symmetry-breaking effects through the appearance of additional angular channels.

%/_/_/_/_/_/_/_/_/_/_/_/_/_/_/_/_/_/_/_/_/_/_/_/_/_/_/_/_/_/_/_/_/_/_/_/_/_/_/_
\section*{Acknowledgments}
%/_/_/_/_/_/_/_/_/_/_/_/_/_/_/_/_/_/_/_/_/_/_/_/_/_/_/_/_/_/_/_/_/_/_/_/_/_/_/_

This work was supported by the JSPS KAKENHI Grants No.~JP21H05019, No.~JP22K04863, No.~JP22H05131 and No.~JP22H05132, and by the JSPS International Joint Research Program JRP- LEAD with UKRI under Grant No.~JPJSJRP20241710. This work was also supported by JST ERATO Grant No.~JPMJER2503.

%/_/_/_/_/_/_/_/_/_/_/_/_/_/_/_/_/_/_/_/_/_/_/_/_/_/_/_/_/_/_/_/_/_/_/_/_/_/_/_
\section*{Data Availability}
%/_/_/_/_/_/_/_/_/_/_/_/_/_/_/_/_/_/_/_/_/_/_/_/_/_/_/_/_/_/_/_/_/_/_/_/_/_/_/_

No data were created or analyzed in this study.

\appendix

\section{Loop phase and chirality}

\subsection{Twist-angle dependence of $\Phi_\triangle$}
\label{appAa}

In the minimal three-wave moir\'e structure, the loop phase is obtained from the phase accumulated along a triangular sequence of scatterings.
Here $\bm r_{ij}$, $\bm r_{jk}$, and $\bm r_{ki}$ denote the directed bond vectors along the three edges of the loop, and the closure condition
$\bm r_{ij}+\bm r_{jk}+\bm r_{ki}=0$
simply expresses that the path returns to its starting point after one circuit around the triangle.
The loop phase is then written as
\begin{equation}
\Phi_\triangle
=
\arg(t_{12}t_{23}t_{31})
=
\bm q_1\!\cdot\!\bm r_{ij}
+
\bm q_2\!\cdot\!\bm r_{jk}
+
\bm q_3\!\cdot\!\bm r_{ki},
\label{eq:app_loop_phase}
\end{equation}
with $|\bm q_1|=|\bm q_2|=|\bm q_3|=q$ and $\bm q_1+\bm q_2+\bm q_3=0$.
The three terms represent the phases accumulated on the three edges, and only their closed sum enters, making $\Phi_\triangle$ gauge invariant.

Introducing the oriented area vector $\bm S_{ijk}=\frac12\,\bm r_{ij}\times\bm r_{jk}$, one may rewrite the phase in the geometric form
\begin{equation}
\Phi_\triangle
=
(\bm q_1\times\bm q_2)\cdot\bm S_{ijk}.
\label{eq:app_loop_area}
\end{equation}
This expression makes the geometric content transparent: the loop phase is controlled by the moir\'e wave-vector area $\bm q_1\times\bm q_2$ and by the real-space area enclosed by the triangular path.
It also makes the pseudoscalar character explicit, since reversing the loop orientation gives $\bm S_{ijk}\to-\bm S_{ijk}$ and hence $\Phi_\triangle\to-\Phi_\triangle$.

Using $q(\theta)=2K\sin(\theta/2)$, Eq.~\eqref{eq:app_loop_area} gives $\Phi_\triangle(\theta)\propto q(\theta)^2 S_{ijk}^{(z)}(\theta)$.
For the minimal triangular loop, this yields the leading dependence
\begin{equation}
\Phi_\triangle(\theta)
=
\mathcal C_\Phi (1-\cos\theta)\sin\theta,
\label{eq:app_phi_theta}
\end{equation}
with $\mathcal C_\Phi$ a dimensionless coefficient set by microscopic details.
Thus $\Phi_\triangle\sim\theta^3$ for small $\theta$, while the same form remains useful as a compact parametrization beyond the strict small-angle limit.

\subsection{Loop-current operator and real-space chirality}
\label{appAb}

Projecting the continuum BM description onto an effective moir\'e lattice gives complex hoppings $t_{ij}$ on the triangular network.
For an effective hopping Hamiltonian
\begin{equation}
\hat H_{\mathrm{eff}}
=
\sum_{\langle ij\rangle}
\Bigl(
t_{ij}\,\hat c_i^\dagger \hat c_j
+\mathrm{h.c.}
\Bigr)
+\cdots,
\label{eq:app_Heff}
\end{equation}
the corresponding bond-current operator follows from the continuity equation, equivalently from $\dot n_i=(i/\hbar)[\hat H_{\mathrm{eff}},\hat n_i]$, as
\begin{equation}
\hat J_{ij}
=
\frac{ie}{\hbar}
\left(
t_{ij}\hat c_i^\dagger \hat c_j
-
t_{ij}^\ast \hat c_j^\dagger \hat c_i
\right),
\qquad
\hat J_{ji}=-\hat J_{ij}.
\label{eq:app_bond_current}
\end{equation}

The same notion of current follows directly from the continuum BM Hamiltonian.
In a real-space second-quantized form,
\begin{eqnarray}
&&\hat H_{\mathrm{BM}}
=
\sum_{\ell=1,2}\int d^2\bm r\;
\hat\psi_\ell^\dagger(\bm r)\,
\hat h_\ell\,
\hat\psi_\ell(\bm r)
\nonumber\\
&&\qquad \qquad+
\int d^2\bm r\;
\Big[
\hat\psi_1^\dagger(\bm r)\,
\hat T(\bm r)\,
\hat\psi_2(\bm r)
+\mathrm{h.c.}
\Big],
\label{eq:app_HBM}
\end{eqnarray}
where $\hat\psi_\ell(\bm r)$ is the sublattice spinor in layer $\ell$ and $\hat h_\ell$ the corresponding Dirac operator.
Defining the layer-resolved charge density
$\hat\rho_\ell(\bm r)=-e\,\hat\psi_\ell^\dagger(\bm r)\hat\psi_\ell(\bm r)$,
the Heisenberg equation gives
\begin{equation}
\partial_t \hat\rho_\ell(\bm r)+\bm\nabla\cdot \hat{\bm j}_\ell(\bm r)
=
\hat{\mathcal I}_\ell(\bm r),
\label{eq:app_continuity}
\end{equation}
where $\hat{\bm j}_\ell$ is the in-plane current and $\hat{\mathcal I}_\ell$ describes interlayer charge transfer induced by tunneling.
Upon projection to localized moir\'e Wannier orbitals $\{|w_i\rangle\}$, one obtains
$t_{ij}=\langle w_i|\hat H_{\mathrm{BM}}|w_j\rangle$,
and Eq.~\eqref{eq:app_bond_current} follows.
This provides the continuum-to-lattice connection underlying the loop-circulation operator $\hat O_{ijk}$.

\subsection{Microscopic $C_3$ covariance of $\mathcal M_{\omega}^{(j)}$}
\label{appAc}

We show that the microscopic channel amplitude
$\mathcal M_{\omega}^{(j)}(\bm k)$ in
Eq.~\eqref{eq:Ma_schematic} is $C_3$-covariant.
In the BM description~\cite{Bistritzer2011PNAS}, the three tunneling operators
$\mathcal T_j$ form a $C_3$ triplet, and the moir\'e wave vectors satisfy
$R_{2\pi/3}\bm q_j=\bm q_{j+1}$.
Accordingly, the three channels are cyclically permuted under
$R_{2\pi/3}$.

Let $U_{C_3}$ be the unitary representation of the threefold rotation.
Since
\begin{equation}
U_{C_3} H(\bm k) U_{C_3}^\dagger = H(R_{2\pi/3}\bm k),
\label{eq:C3_H_appendix}
\end{equation}
the Bloch states may be chosen as
\begin{equation}
U_{C_3}|u_{\bm k,n}\rangle
=
e^{i\gamma_n(\bm k)}
|u_{R_{2\pi/3}\bm k,n}\rangle ,
\label{eq:C3_bloch_appendix}
\end{equation}
with the initial state understood as the corresponding band.
The operators transform as
\begin{equation}
U_{C_3} v_a U_{C_3}^\dagger = (R_{2\pi/3})_{ab} v_b,
\qquad
U_{C_3}\mathcal T_j U_{C_3}^\dagger = \mathcal T_{j+1},
\label{eq:C3_ops_appendix}
\end{equation}
and the band energies obey
$\varepsilon_{R_{2\pi/3}\bm k,n}=\varepsilon_{\bm k,n}$.
Substituting Eqs.~\eqref{eq:C3_bloch_appendix} and
\eqref{eq:C3_ops_appendix} into Eq.~\eqref{eq:Ma_schematic}, together with
$R_{2\pi/3}\bm q_j=\bm q_{j+1}$, gives $\mathcal M_{\omega}^{(j)}(R_{2\pi/3}\bm k)=\mathcal M_{\omega}^{(j+1)}(\bm k)$, up to a basis-dependent phase that cancels in gauge-invariant combinations
such as $\mathcal M_{\omega}^{(1)}\mathcal M_{\omega}^{(2)}\mathcal M_{\omega}^{(3)}$. Thus the three amplitudes form a single $C_3$ channel triplet.

\end{document}